\documentclass[12pt]{spieman}  

\usepackage{amsmath,amsfonts,amssymb,amsthm}
\usepackage{graphicx}
\usepackage{setspace}
\usepackage{tocloft}
\usepackage{lineno}
\usepackage{tabularx}
\usepackage{geometry,latexsym}

\usepackage[symbol]{footmisc}

\title{Implicit Electric Field Conjugation Through a Single-mode Fiber}

\author[a, c, *]{Joshua Liberman}
\author[a,b]{Jorge Llop-Sayson}
\author[a]{Arielle Bertrou-Cantou}
\author[a,b]{Dimitri Mawet}
\author[a]{Niyati Desai}
\author[d]{Sebastiaan Y Haffert}
\author[b]{A J Eldorado Riggs}
\affil[a]{California Institute of Technology, Pasadena, California, United States}
\affil[b]{Jet Propulsion Laboratory, California Institute of Technology, Pasadena, California, United States}
\affil[c]{James C. Wyant College of Optical Sciences, University of Arizona,
Meinel Building 1630 E. University Blvd., Tucson, AZ. 85721}
\affil[d]{Steward Observatory, University of Arizona, 933 N Cherry Ave, Tucson, AZ, USA 85719}

\cftpagenumbersoff{figure}
\cftpagenumbersoff{table} 
\begin{document} 
\maketitle

\begin{abstract}
Connecting a coronagraph instrument to a spectrograph via a single-mode optical fiber is a promising technique for characterizing the atmospheres of exoplanets with ground and space-based telescopes. However, due to the small separation and extreme flux ratio between planets and their host stars, instrument sensitivity will be limited by residual starlight leaking into the fiber. To minimize stellar leakage, we must control the electric field at the fiber input. Implicit electric field conjugation (iEFC) is a model-independent wavefront control technique in contrast with classical electric field conjugation (EFC) which requires a detailed optical model of the system. We present here the concept of an iEFC-based wavefront control algorithm to improve stellar rejection through a single-mode fiber. As opposed to image-based iEFC which relies on minimizing intensity in a dark hole region, our approach aims to minimize the amount of residual starlight coupling into a single-mode fiber. We present broadband simulation results demonstrating a normalized intensity $\ge 10^{-10}$ for both fiber-based EFC and iEFC. We find that both control algorithms exhibit similar performance for the low wavefront error (WFE) case, however, iEFC outperforms EFC by $\approx$ 100x in the high WFE regime. Having no need for an optical model, this fiber-based approach offers a promising alternative to EFC for ground and space-based telescope missions, particularly in the presence of residual WFE. 
\end{abstract}


{\noindent \footnotesize\textbf{*}Joshua Liberman,  \linkable{jliberman@arizona.edu} }

\begin{spacing}{2}   

\section{Introduction}
\label{sect:intro}  

{Detection and characterization of Earth-like planets in the habitable zone is challenging due to the small angular separation and high contrast between exoplanets and their host stars. High contrast imaging (HCI) systems aim to suppress diffracted starlight, revealing a faint planet signal.} {F}uture space-based telescope missions, such as the Habitable Worlds Observatory, (see also the Habitable Exoplanet Observatory (HabEx) and the Large UV/Optical/IR Surveyor (LUVOIR) concepts\cite{luvoir2019, habex2020}) {will focus on the detection and characterization of exo-Earths around Sun-like stars.} {S}uch missions will require raw contrasts of $\simeq 10^{-10}$ at a close angular separation of $\approx 0.1"${, which induce strict wavefront quality and stability requirements\cite{Currie2022}. The next generation of ground-based extremely large telescopes (ELTs) will target exo-Earths around low-mass stars. While the raw contrast requirements are less stringent ($\simeq10^{-5}$ at $\approx 0.015"$), ELTs must also deal with turbulent residuals appearing downstream of the extreme adaptive optics system \cite{kasper2021}.}

{Assuming the planet position is well constrained, the} coronagraph {can be connected} to a medium-to-high resolution spectrograph via an optical fiber {to} enhance rejection of starlight at small inner working angles\cite{snellen2015, mawet2017}. Furthermore, {high dispersion coronagraphy (HDC)\cite{snellen2015,mawet2017}} can be used to measure planetary orbital velocities and atmospheric compositions{\cite{Wang2021}}. Modern {ground-based} instruments such as the Keck Planet Imager and Characterizer (KPIC) and the Subaru Telescope's Rigorous Exoplanetary Atmosphere Characterization with High dispersion coronography (REACH) demonstrate the feasibility of HDC on diffraction-limited telescopes by coupling planet light into a single-mode fiber (SMF)\cite{delorme2021, kotani2020}. {Additionally, next generation ground-based HDCs, such as the Planetary Systems Imager (PSI) on the Thirty Meter Telescope (TMT) are in development and future space-based HDC concepts are being considered\cite{fitzgerald2022, wang2017}.} 

The SMF further rejects starlight due to its mode selectivity\cite{LlopSayson2019, Coker2019}. However, despite the increased stellar {suppression}, the signal-to-noise ratio (SNR) of the planet spectrum remains limited due to residual starlight leaking into the fiber. This leakage results in excess photon noise and contaminates the planet signal\cite{LlopSayson2021, Xin2023}.

{Wavefront control (WFC) is aimed at suppressing stellar speckles to help recover the companion's signal. However, before WFC can be performed, the electric field must be measured in the focal plane. Two common methods for recovering the focal plane electric field are pair-wise probing (PWP), which utilizes phase diversity and the self coherent camera (SCC), where a pinhole is added in the Lyot Stop plane, creating fringes in the image plane that spatially encode the stellar speckles\cite{giveon2007, baudoz2006, potier2020}.} In {WFC}, one or two deformable mirrors (DM{s}) modify the incoming electric field to create a dark hole (DH) region, free of speckles\cite{LlopSayson2019}. While many wavefront control techniques exist, Electric Field Conjugation (EFC) is often used due to its superior performance relative to most other controllers\cite{groff2016}. EFC is the primary controller for the Nancy Grace Roman Space Telescope \cite{kasdin2020}. {Additionally, EFC has been implemented on{-sky with} the VLT/SPHERE instrument\cite{Potier-2022}.} EFC involves minimizing the focal plane electric field, using an optimal DM shape. The DM pattern is derived from a detailed instrument model\cite{giveon2007}. 

The model-dependency of EFC limits its feasibility on-sky, as any telescope system instabilities will reduce the algorithm's effectiveness \cite{Haffert2023, matthews2017, Potier-2022}. {Moreover, space-based telescopes too suffer from model uncertainties due to optical misalignments on-orbit caused by changing thermal gradients\cite{derby2023}.} Implicit EFC (iEFC) utilizes a linear response between the DM and modulated intensity measurements, such that the reconstructed electric field is no longer computed explicitly\cite{Haffert2023, ahn_guyon_2023, ruffio_2022}. {iEFC is not the only measurement-based WFC technique. On the In Air Contrast Testbed (IACT) at NASA JPL, SCC was combined with EFC to create another empirical WFC method. A comparison between iEFC, SCC+EFC, and PWP+EFC showed similar performance ($\simeq 10^{-8}$) with a Vector Vortex Coronagraph (VVC)\cite{desai2023}.} In this work, we propose an algorithm based on iEFC to minimize the speckles coupling into a SMF {for both ground- and space-based systems}. In Section~\ref{sect:theory}, we derive the equations for fiber-based iEFC. In Section~\ref{sect:sims}, we describe our simulation results, and in Section~\ref{sect:Discussion}, we discuss and conclude our work.

\section{Theory}
\label{sect:theory}

\subsection{EFC With a Single-Mode Fiber}
\label{subsect:efc-smf}
{
We begin by defining a complex input electric field
\begin{equation}
\label{input-e-field}
    E_{\mathrm{in}} = E_{\mathrm{0}} + \Delta E
\end{equation}
where $E_{\mathrm{0}}$ represents the diffraction-limited electric field and $\Delta E$ represents the change in the electric field induced by aberrations.
We can express our final focal plane electric field as 
\begin{equation}
    \label{efield-fp}
    E_{\mathrm{FP}} = C\left \{E_{\mathrm{in}} \right \} = C\left \{ E_{\mathrm{0}} \right \} + C\left \{ \Delta E \right \}
\end{equation}
where $C$ represents a coronagraph operator that propagates $E_{\mathrm{in}}$ through a coronagraphic optical system via a series of Fourier transforms. {Following the methodology of Ref.~\citenum{LlopSayson2019}}, we may express our output electric field propagated through the SMF as
\begin{equation}
    \label{efield-smf}
    E_{\mathrm{SMF}} = D\left \{E_{\mathrm{FP}} \right \}
\end{equation}. 
$D$ represents an operator for the overlap integral of the focal plane electric field at the fiber input multiplied by the fiber's fundamental mode. Operating on $E_{\mathrm{FP}}$, we obtain 
\begin{equation}
    \begin{gathered}
        \label{Esmf-fp}
           D \left \{E_{\mathrm{FP}} \right \} = E_{\mathrm{SMF}} = \int E_{\mathrm{FP}} \Psi^{\dagger}_{\mathrm{SMF}} \mathrm{d}a  = \overline{E}_{\mathrm{0}} + \Delta  \overline{E}
    \end{gathered}
\end{equation}
where $\Psi^{\dagger}_{\mathrm{SMF}}$ denotes the complex conjugate of the fiber mode shape and $\mathrm{d}a$ denotes the differential area element in the image plane. $\overline{E}_\mathrm{0}$ and $\Delta \overline{E}$ denote the propagation of $E_{\mathrm{0}}$ and $\Delta E$ through the SMF in the focal plane.
Expressing Equation~\eqref{Esmf-fp} as an intensity, we obtain
\begin{equation}
    \begin{gathered}
        \label{intensity-fp-squared}
        \left | E_{\mathrm{SMF}} \right|^{2} = \left | \overline{E}_{\mathrm{0}} \right |^{2} + \left | \Delta \overline{E} \right |^{2} + 2\Re\left \{ \overline{E}_{\mathrm{0}}  \Delta \overline{E}^{\dagger} \right \}
    \end{gathered}
\end{equation}
where $\Delta \overline{E}^{\dagger}$ represents the complex conjugate of $\Delta \overline{E}$.
EFC requires a detailed estimate of the electric field, which we can obtain via PWP. \footnote{Detailed descriptions of the PWP method for EFC can be found in Ref.~\citenum{GiveOn2011} and Ref.~\citenum{groff2016}, while additional information on PWP through a SMF can be found in Ref.~\citenum{LlopSayson2019}.} $\Delta \overline{E}$ can be further decomposed into 
\begin{equation}
    \Delta \overline{E} = \Delta \overline{E}_{\mathrm{DM}} + \Delta \overline{E}_{\mathrm{ab}}
\end{equation}
where $\Delta \overline{E}_{\mathrm{DM}}$ denotes the change in electric field induced by a sine wave, or probe, applied to the DM surface and $\Delta \overline{E}_{\mathrm{ab}}$ denotes the non-common path aberrations through the SMF in the final focal plane. Now, we may linearize the problem by assuming that $\Delta \overline{E} << 1$ and $\Delta \overline{E}_{\mathrm{0}} = 0$.\footnote{Note that the method of EFC breaks down when this linear assumption no longer holds. This is particularly relevant in on-sky applications when $\Delta \overline{E}$ is no longer $<<1$. By setting $\Delta \overline{E}_{\mathrm{0}} = 0$, we assume a perfect coronagraph which is a reasonable approximation for the VVC.} We can now create pairs of intensity images by applying positive and negative sine probes ($\overline{E}_{\mathrm{DM}}$ and $-\overline{E}_{\mathrm{DM}}$) on the DM. Our image pairs can be written as
\begin{equation}
    \begin{gathered}
        \label{intensity-plus}
        I^{+}_{\mathrm{FP}} =  \left | \Delta \overline{E}^{+} \right|^{2} = \left | \Delta \overline{E}_{\mathrm{DM}} \right |^{2} + \left | \Delta \overline{E}_{\mathrm{ab}} \right |^{2} + 2\Re\left \{ \Delta \overline{E}_{\mathrm{DM}} \Delta \overline{E}_{\mathrm{ab}}^{\dagger} \right \}
    \end{gathered}
\end{equation}
and
\begin{equation}
    \begin{gathered}
        \label{intensity-minus}
      I^{-}_{\mathrm{FP}} =  \left | \Delta \overline{E}^{+} \right|^{2} = \left | \Delta \overline{E}_{\mathrm{DM}} \right |^{2} + \left | \Delta \overline{E}_{\mathrm{ab}} \right |^{2} - 2\Re\left \{ \Delta \overline{E}_{\mathrm{DM}} \Delta \overline{E}_{\mathrm{ab}}^{\dagger} \right \}
    \end{gathered}
\end{equation}.
Subtracting $ I^{-}_{\mathrm{FP}}$ from $ I^{+}_{\mathrm{FP}}$ and re-writing the result in matrix form, we obtain
\begin{equation}
    \begin{gathered}
        \label{pwp}
     \begin{bmatrix}
    \Delta I_{\mathrm{1}}\\ 
    \vdots \\ 
    \Delta I_{n}
    \end{bmatrix} = 4 \begin{bmatrix}
    \Re \left \{ \Delta \overline{E}_{\mathrm{DM}} \right \} & \Im \left \{ \Delta \overline{E}_{\mathrm{DM}} \right \} \\
    \vdots & \vdots \\
    \Re \left \{ \Delta \overline{E}_{\mathrm{DM}_{n}} \right \} & \Im \left \{ \Delta \overline{E}_{\mathrm{DM}_{n}} \right \}
    \end{bmatrix} \begin{bmatrix}
    \Re \left \{\Delta \overline{E}_{\mathrm{ab}} \right \} \\ \Im \left \{\Delta \overline{E}_{\mathrm{DM}} \right \}
    \end{bmatrix}
    \end{gathered}
\end{equation}
for $n$ pairwise probes.
Letting 
\begin{equation}
    c =  \begin{bmatrix}
    \Delta I_{1}\\ 
    \vdots \\ 
    \Delta I_{n}
    \end{bmatrix} 
\end{equation},
\begin{equation}
    \textbf{G} = 4 \begin{bmatrix}
    \Re \left \{ \Delta \overline{E}_{\mathrm{DM}} \right \} & \Im \left \{ \Delta \overline{E}_{\mathrm{DM}} \right \} \\
    \vdots & \vdots \\
    \Re \left \{ \Delta \overline{E}_{\mathrm{DM}_{n}} \right \} & \Im \left \{ \Delta \overline{E}_{\mathrm{DM}_{n}} \right \}
    \end{bmatrix}
\end{equation}
, and
\begin{equation}
    a = \begin{bmatrix}
    \Re \left \{\Delta \overline{E}_{\mathrm{ab}} \right \} \\ \Im \left \{\Delta \overline{E}_{\mathrm{ab}} \right \}
    \end{bmatrix}
\end{equation}
, we may construct an expression to estimate $\Delta \overline{E}_{ab}$, given by
\begin{equation}
    c = \textbf{G}a
\end{equation}. 
With $\hat{a} = \textbf{G}^{-1}c$ representing our WF estimate at each control iteration, we can now solve for a DM solution that will minimize our estimate, thus minimizing the intensity through the SMF. Our least squares solution is given as
\begin{equation}
    u = \mathrm{argmin}  \left |a + \textbf{G}u \right|^{2}
\end{equation},
where $\textbf{G}$ represents the Jacobian matrix relating actuator pokes to their effects on the overlap integral and $u$ represents the DM solution that minimizes starlight coupling into the SMF. These effects are determined from a model of the optical system.}

\subsection{iEFC With a Single-Mode Fiber}
\label{subsec:iEFC-smf}
{Let us now define a basis of Fourier modes, or sine waves, on the DM. Let $\Delta \overline{E}_{M}$ represent the change in the focal plane electric field through a SMF after applying a Fourier mode. We may again collect pairwise intensity images by exciting positive and negative modes. The relationship between the excited mode states and the pairwise intensity images can be linearly represented as
\begin{equation}
    \begin{gathered}
        \label{iefc-lin-relation}
     \begin{bmatrix}
    \Delta I_{1}\\ 
    \vdots \\ 
    \Delta I_{n}
    \end{bmatrix} = 4 \begin{bmatrix}
    \Re \left \{ \Delta \overline{E}_{\mathrm{M}_{\mathrm{1}}} \right \} & \Im \left \{ \Delta \overline{E}_{\mathrm{M}_{\mathrm{1}}} \right \} \\
    \vdots & \vdots \\
    \Re \left \{ \Delta \overline{E}_{\mathrm{M}_{\mathrm{n}}} \right \} & \Im \left \{ \Delta \overline{E}_{\mathrm{M}_{\mathrm{n}}} \right \}
    \end{bmatrix} \begin{bmatrix}
    \Re \left \{\Delta \overline{E}_{\mathrm{ab}} \right \} \\ \Im \left \{\Delta \overline{E}_{\mathrm{ab}} \right \}
    \end{bmatrix}
    \end{gathered}
\end{equation}.
Alternatively, we may write Equation~\eqref{iefc-lin-relation} as 
\begin{equation}
    c = \textbf{M}a
\end{equation}
where $\textbf{M}$ denotes our matrix relating the effect of each Fourier mode on the overlap integral. $\textbf{M}$ and $\textbf{G}$ may now be combined to form
\begin{equation}
    \label{iefc-inter-mat}
    \Delta I = \textbf{M} \textbf{G} u = \textbf{Z} u
\end{equation}
where $\textbf{Z}$ describes a response matrix relating our pairwise probe images to our pairwise mode images. We may now compute a DM solution so as to minimize the overlap integral,
\begin{equation}
    u = \mathrm{argmin}  \left |c + \textbf{Z}u \right|^{2}
\end{equation}. Thus, we have iteratively obtained a measurement-based DM solution that suppresses starlight through the SMF.}
\section{Simulations}
\label{sect:sims}
To assess performance of iEFC through a {SMF}, we developed an end-to-end optical simulation based on the current layout of Caltech's High Contrast Spectroscopy Testbed (HCST)\cite{Bertrou2023SPIE}. 

Our simulation was constructed with the Fast Linearized Coronagraph Optimizer (FALCO) package within MATLAB, which provides realistic optical propagations for coronagraphs\cite{falco2018}. {All simulation trials were computed with a central wavelength of $\lambda_{0} = 750~nm$.}We simulated a 55 segment off-axis telescope pupil, based on the LUVOIR-B design, with a circularly clipped outer edge. The circumscribed telescope diameter is 7.989 meters with each segment measuring 0.955m flat-to-flat. The segment gaps are $< 0.1\%$ of the pupil diameter\cite{potier-spie-2022}. We also simulated a Lyot stop with a diameter of {82\% of the incoming beam diameter}, a charge 6 VVC, a 34x34 actuator Boston Micromachines DM, and a single-mode fiber with a mode field radius of {$0.507~\lambda_{0} /D$}. 

{In both EFC and iEFC simulations, we used 2 DM probes. The first DM probe is a sine wave, generated by summing over all of the even Fourier modes in our defined control region. The second DM probe is a cosine wave, generated by summing over all of the odd Fourier modes within our control region. By creating both sine and cosine probes, we capture both the real and imaginary components of the electric field.}

Realistic speckles were generated using a phase map spanning low-high spatial frequencies (nolls 4-100) with an RMS wavefront error (WFE) of 10~nm--consistent with the aberrations we observe on our testbed\cite{hcst2019}. We neglect the impact of segment phasing errors for our study, instead focusing on WFE that spans the pupil as a whole.

\begin{figure}
\begin{center}
\begin{tabular}{c}
\includegraphics[height=5.5cm]{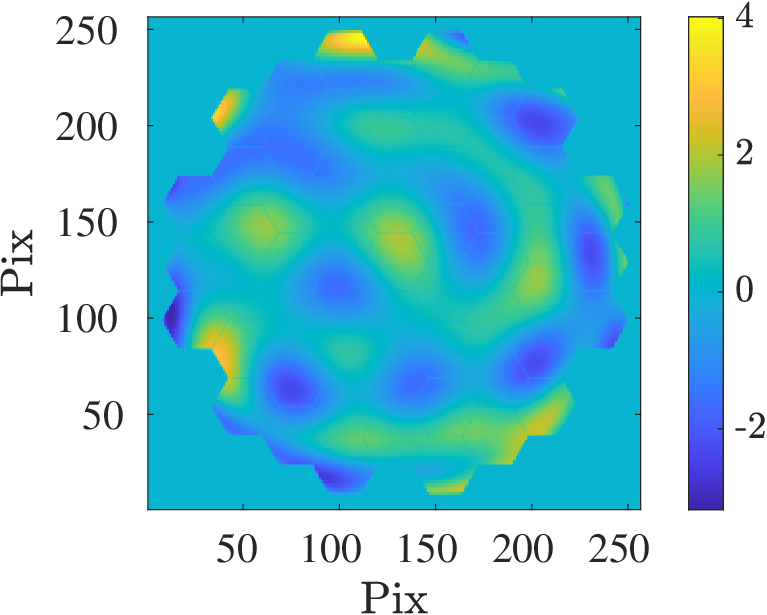} 
\end{tabular}
\end{center}
\caption 
{ \label{fig:phase-map}
The LUVOIR-B phase map used in HCST simulations. {The intensity is defined in units of nanometers.}}
\end{figure} 

In defining contrast performance, we adopt the formalism of Ref.~\citenum{LlopSayson2019}. We define SMF normalized intensity (NI) as the measured power at the output of the SMF divided by measured intensity at the output of the SMF centered on the non-coronagraphic stellar PSF.

\begin{figure}
\begin{center}
\includegraphics[height=5cm]{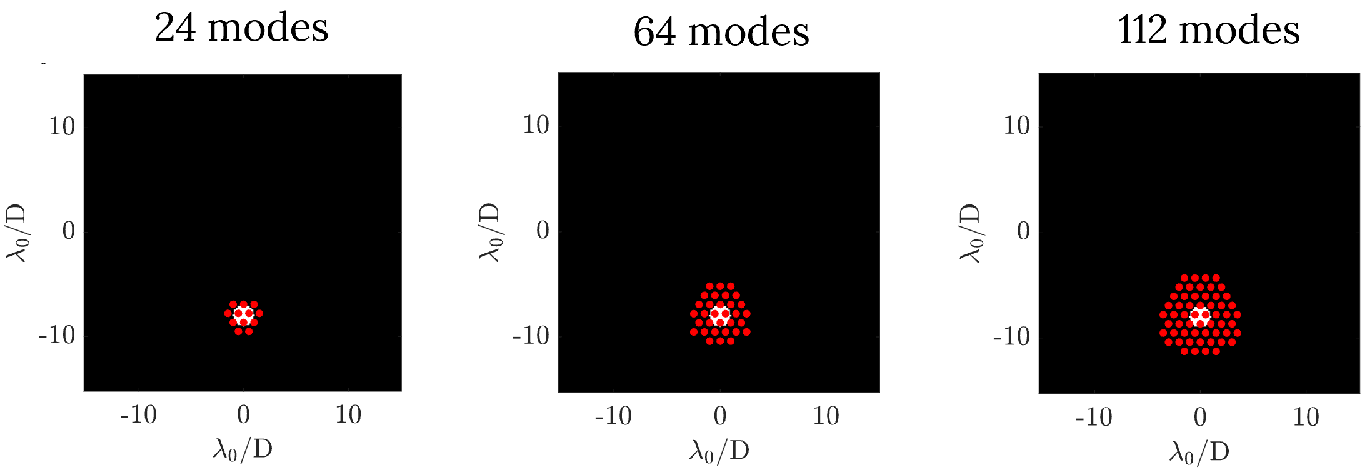}
\end{center}
\caption 
{ \label{fig:mode-comp}
Three different Fourier mode configurations overlaid on a 2-D projection of the fiber's fundamental mode at (0, -8) {$\lambda_{0} /D$. The red dots represent pairs of 12, 32, and 56 modes, respectively.}}

\end{figure}

We first simulate fiber-based iEFC and EFC at 2 different locations within our image, using a circular control region of 1~{$\lambda_{0} /D$} to match the dimensions of our fiber and a fixed configuration of 24 Fourier modes. {This mode configuration was chosen so as to minimize edge effects. Four Fourier modes fall inside the nulled region where the fiber is centered (Fig.~\ref{fig:mode-comp}), however, we choose to oversize this area to 12 Fourier mode pairs, or 24 modes, in order to control spatial frequencies at the border of the dark hole.}

\begin{figure}
\begin{center}
\begin{tabular}{c}
\includegraphics[height=6.5cm]{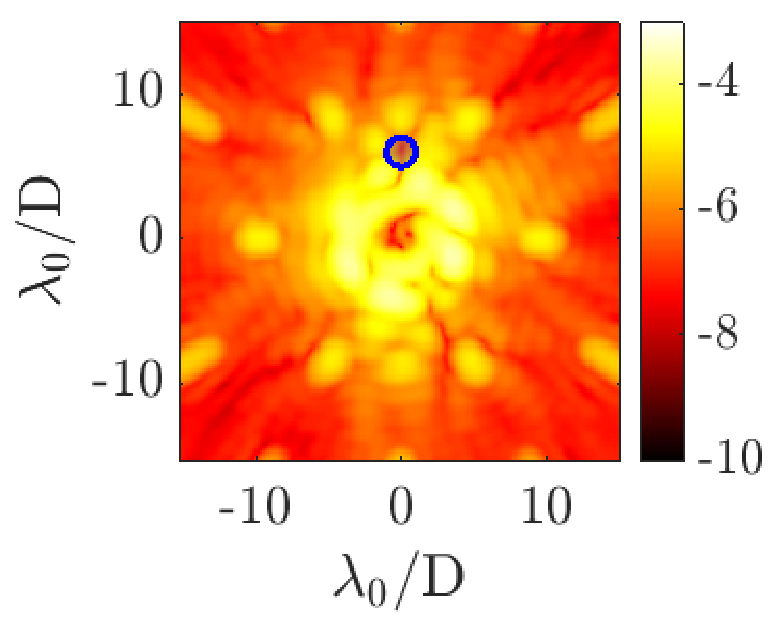}
\end{tabular}
\end{center}
\caption 
{ \label{fig:psf}
A simulated stellar PSF after correction with fiber-based iEFC. The {blue} circle marks the position of the fiber at (0, 6) {$\lambda_{0} / D$. The colorbar denotes normalized intensity in log scale.}} 
\end{figure}
\begin{figure}
\begin{center}
\begin{tabular}{c}
\includegraphics[height=7.5cm]{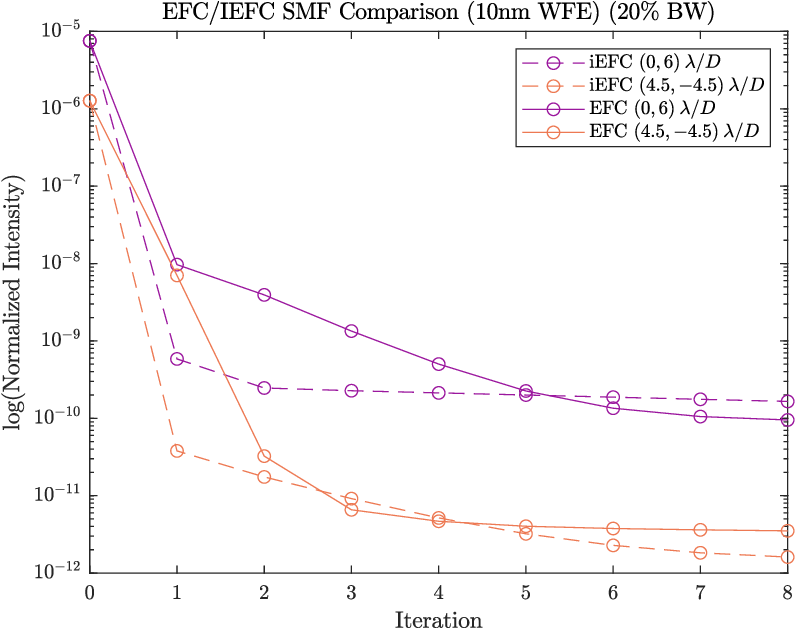}
\end{tabular}
\end{center}
\caption 
{ \label{fig:iefc_efc_bb20}
Contrast performance of EFC versus iEFC through a SMF at 2 different fiber locations over a bandwidth of $\Delta \lambda / { \lambda_{0}} = 20\%$.} 
\end{figure}

A simulated stellar PSF is shown in Fig.~\ref{fig:psf}. Note that fiber-based iEFC will not necessarily create a DH at a given fiber position, as it is minimizing the overlap integral rather than the electric field \cite{LlopSayson2019}. In Fig.~\ref{fig:iefc_efc_bb20}, we compare performance of iEFC to that of EFC in polychromatic light with a {$\Delta \lambda / \lambda_{0} = 20\%$} bandwidth{,} 9~sub-bandpasses{, and a circular control region with a diameter of 1 $\lambda_{0} / D$}.

All solutions converge on a NI of $\approx 10^{-10}$ or lower {in fewer than 8 iterations}. {Additionally, we} observe similar performance between EFC and iEFC {in simulations}. {The similarities in achieved contrast between EFC and iEFC are} consistent with {Ref~\citenum{Haffert2023}'s conclusions. We also} observe some variation in the NI as a function of the fiber position. This is to be expected, as the phase pattern of the electric field along the fiber tip, the orientation of the Fourier modes with respect to the fiber, and the WFE all result in variation of the NI.

\begin{figure}
\begin{center}
\begin{tabular}{c}
\includegraphics[height=7.5cm]{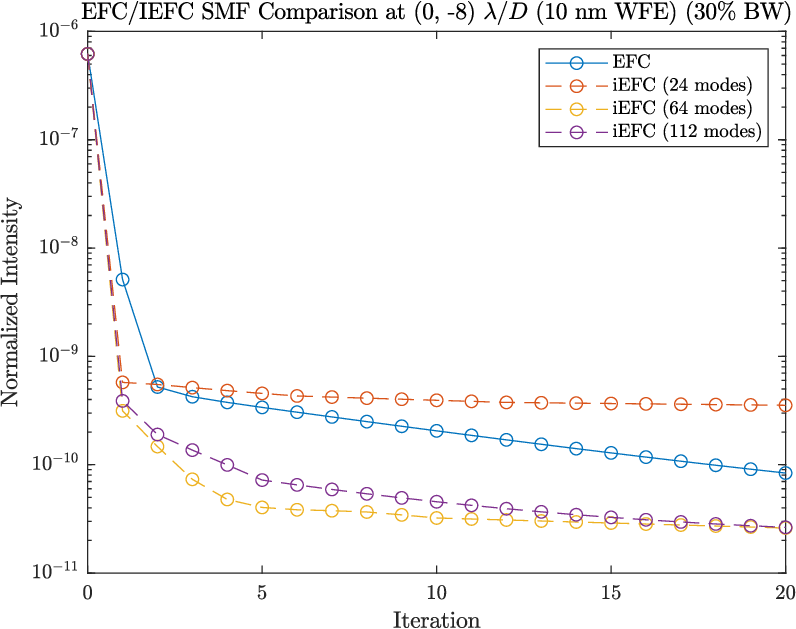}
\end{tabular}
\end{center}
\caption 
{ \label{fig:iefc-efc-mode-comp}
EFC versus iEFC contrast for a {$\Delta \lambda / \lambda_{0} = 30\%$} bandwidth. iEFC performance is evaluated for three different mode configurations.} 
\end{figure}

{The segmented LUVOIR-B pupil used in our simulations may produce non-linear cross talk between Fourier modes. Hence, more modes would be necessary if certain modes contribute more power to the WFC solution than others.}
{To examine the impacts of non-linear cross talk, we} simulate three different Fourier mode configurations at a fixed fiber location of (0, -8) {$\lambda_{0} /D$}. Fig~\ref{fig:mode-comp} shows the Fourier modes overlaid on a 2-D projection of the fiber's fundamental mode. {The modes are spaced at intervals of 1 cycle/pupil.}
In Fig.~\ref{fig:iefc-efc-mode-comp}, we compare the NI of EFC to that of iEFC for a varying number of Fourier modes and a bandwidth of {$\Delta \lambda / \lambda_{0} = 30\%$} over 13 sub-bandpasses. {Over 20 iterations, }EFC and iEFC again exhibit similar similar performance through the fiber, however, iEFC achieves higher contrasts (NI$\approx 10^{-11}$) as we increase the number of modes. {iEFC with 112 modes is slower than that of the 64 mode configuration, yet both solutions converge on an NI of $\approx 10^{-11}$. The results of Fig.~\ref{fig:iefc-efc-mode-comp} demonstrate that past 64 modes, there is no benefit to increasing the number of modes, as the system edge effects become negligible. However, additional work is needed to identify the optimal mode configuration for the iEFC controller. With a more optimal mode selection, we may be able to reduce the number of modes needed for the Jacobian measurement, thus decreasing the measurement time.}

\begin{figure}
\begin{center}
\begin{tabular}{c}
\includegraphics[height=7.5cm]{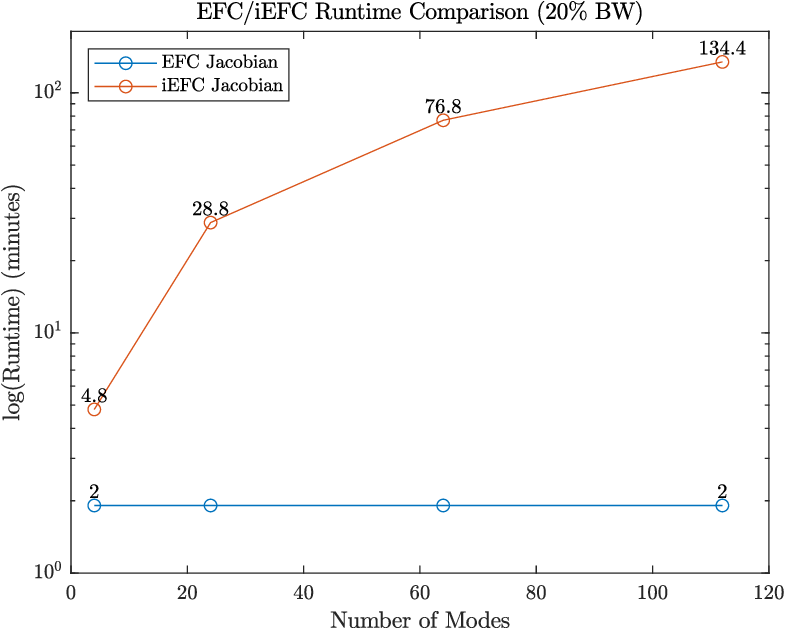}
\end{tabular}
\end{center}
\caption 
{ \label{fig:jacobian-timing}
{Estimated }time required to construct the Jacobian for iEFC vs EFC {in real observations}, assuming an arbitrary integration time of 1 second and a bandwidth of {$\Delta \lambda / \lambda_{0} = 20\%$.} In iEFC, the Jacobian measurement time scales with the number of modes. In the case of EFC, no images are required to compute the Jacobian--only an optical model propagation is necessary.} 
\end{figure}

While we find that iEFC contrast improves with the number of Fourier modes, it is important to consider the increased Jacobian measurement time for iEFC in this analysis. In Fig.~\ref{fig:jacobian-timing}, we {estimate} the time to construct the Jacobian in iEFC vs EFC {for real observations. We assume an arbitrary $\Delta \lambda / \lambda_{0} = 20\%$} bandwidth with 9 sub-bandpasses. {We also assume that the detector readout time is negligible. F}or EFC, the Jacobian construction time is obtained directly from our FALCO simulation. In iEFC, we estimate the Jacobian construction time, assuming 1-second integration{s}.  

We observe that as the number of modes are increased, the time to construct the iEFC measurement Jacobian {scales} significantly. On the other hand, the EFC Jacobian construction time remains fixed, as it is computed from a model with a fixed number of modes. The measurement time of the iEFC response matrix can be prohibitive for certain operations in which the time scale of the WF variations is similar to that of the response matrix. However, this should not pose a constraint when the coronagraph instrument is stable enough such that a single response matrix can reflect the status of the WFE over an extended period of time. Future space-based imaging missions, in particular, will require such stability as well as a robust wavefront sensing and control loop to reach at least $10^{-10}$ contrast\cite{habex2020}. {Thus, the choice of mode configuration becomes a compromise between the desired contrast level and the Jacobian computation time.} Note that, for EFC, the issue of long Jacobian computation times is not relevant, as WFE is not included within the optical model. Also note that Fig.~\ref{fig:jacobian-timing} does not account for the additional time required for wavefront sensing (WFS) with EFC. The estimated EFC WFS time is $\approx 54$ seconds which is shorter than the respective Jacobian computation time.

\begin{figure}
\begin{center}
\begin{tabular}{c}
\includegraphics[height=7.5cm]{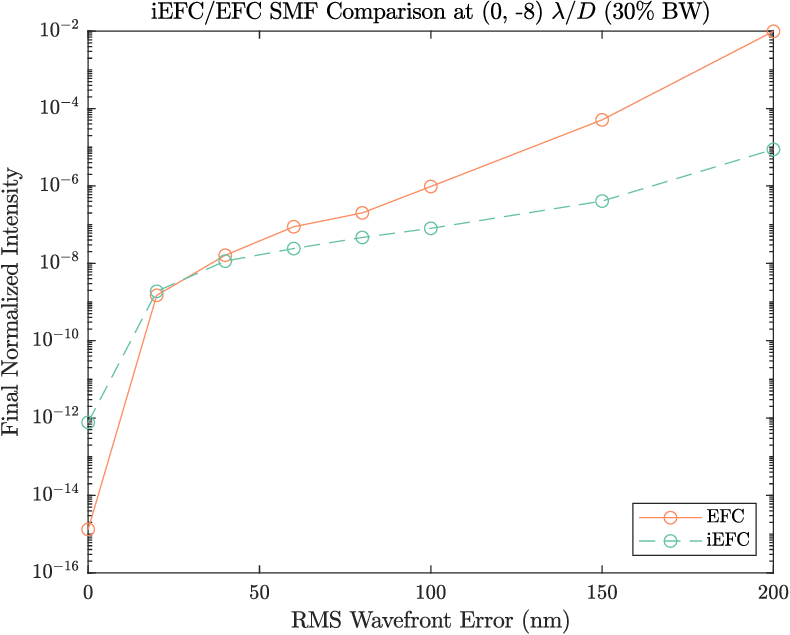}
\end{tabular}
\end{center}
\caption 
{ \label{fig:model-errors}
Final normalized intensity over a range of injected model errors for a {$\Delta \lambda / \lambda_{0} = 30\%$} bandwidth.
} 
\end{figure}

{When implementing EFC and iEFC on ground-based instruments, the model errors are much larger than those of space-based instruments due to the presence of a turbulent atmosphere.} In Fig.~\ref{fig:model-errors}, we examine how EFC and iEFC respond to model errors within our simulation over a {$\Delta \lambda / \lambda_{0} = 30\%$ bandwidth} and 13 sub-bandpasses. {These errors are roughly consistent with what we would observe on ground-based telescopes\cite{gilles2008}.} To generate model errors, we use the phase map shown in Fig.~\ref{fig:phase-map} and vary the amplitude of RMS WFE. For iEFC, we use a fixed configuration of 24 Fourier modes. Both algorithms converge on a solution within 9 iterations until $\approx 150$ nm RMS WFE at which point EFC begins to diverge. Furthermore, as the WFE amplitude is increased, the contrast achieved with EFC decreases. This is consistent with what we expect as EFC computes a control solution from an ideal optical model. However, for iEFC in the high WFE regime, the solution remains fairly constant, with iEFC showing a $\approx$100x contrast improvement over EFC with 150~nm RMS WFE.

\section{Discussion and Conclusion}
\label{sect:Discussion}
We describe a new method of focal plane wavefront control through a SMF {that is applicable to both ground- and space-based telescopes}. The model-free approach in iEFC combined with the mode selectivity of a SMF provides a simple method of wavefront control at higher contrasts than that of conventional EFC or iEFC\cite{Haffert2023, LlopSayson2019}. 

In simulations, we consistently demonstrate broadband contrasts of at least $10^{-10}$ with a charge 6 VVC {in the low model error regime and contrasts of $\approx 10^{-9}$ in the high model error regime.} {We} observe similar performance between EFC and iEFC through the fiber {when model errors are small, but the performance of EFC degrades when simulating RMS WFE that is consistent with what we would see on a ground-based telescope}. {This is a particularly notable result, as EFC performance on ground-based telescopes may be limited by the model-based Jacobian solution\cite{matthews2017}. Thus, high contrasts with fiber-based iEFC may be achievable on current and future ground-based fiber-fed spectrographs such as the Keck-KPIC, Keck-HISPEC, and TMT-MODHIS instruments \cite{delorme2021, hispec-mawet-2022}. iEFC is also a promising control algorithm for space-based telescopes, as it eliminates the challenges associated with modelling on-orbit optical degradations\cite{derby2023}.} 

{We observe that a greater number of modes correlates with a higher contrast in iEFC through a SMF up to 64 modes. Beyond this modal configuration, the edge effects become negligible, as our solutions converge on an NI of $\approx 10^{-11}$. The improved performance with 64 Fourier modes could result from a less-than ideal modal selection. The effectiveness of individual modes may be reduced, as there are no trivial modes that would be orthogonal for a segmented pupil. This principle can be extended to EFC as well. Selecting the orthogonal modes could also make it more difficult for EFC and iEFC to converge. Through our runtime comparisons of the Jacobian between EFC and iEFC, we observe that the choice of mode configuration is a trade-off between the desired contrast level and the Jacobian construction time. This result is particularly notable for future space-based imaging systems, as the coronagraph instrument will need to remain well-aligned over the duration of the WFC calibrations.} 

Future work will be needed to identify the optimal number of modes in our iEFC control implementation. {By optimizing the modal selection within iEFC, we may be able to reduce the number of modes needed to construct the measurement Jacobian, thus decreasing the time required for calibrations. We will also investigate how changing the Fourier mode spacing might affect our WFC solution. Additionally, we plan on developing a more robust simulation for iEFC on ground-based telescopes. If the interaction matrix is measured on-sky, we must account for effects such as temporal wavefront errors occurring over our calibration timescale. Such errors may impact our final WFC solution.} Furthermore, we plan on conducting fiber-based iEFC experiments using the fiber injection unit on HCST. Due to its model-free approach and ease of implementation, iEFC through a SMF is a promising path forward for achieving high contrasts in future ground and space-based missions. 

\subsection*{Disclosures}
The authors have no relevant financial interests in the manuscript and no other potential conflicts
of interest to disclose.

\subsection* {Acknowledgments}
{We thank the anonymous referees whose thorough comments improved the quality of this paper. Additionally, we thank Kian Milani for his contributions to the PWP theory in EFC.}
This work was supported by NASA SAT under award number
80NSSC20K0624. This work has been adapted from a prior SPIE proceedings\cite{Liberman2023SPIE}. Joshua Liberman is a member of UCWAZ Local 7065.

\subsection* {Data, Materials, and Code Availability} 
The code and data used in this work can be found \href{https://github.com/jlibermann/falco-matlab}{here}.


\bibliography{report}   
\bibliographystyle{spiejour}   


\vspace{2ex}\noindent\textbf{Joshua Liberman} is a 1st-year Optical Sciences PhD student at the University of Arizona. Previously, he served as a research technician in the Exoplanet Technology lab at Caltech where he assisted with the commissioning of the Keck Planet Imager and Characterizer instrument and developed wavefront control techniques for Caltech's high contrast testbed. He received a Bachelor's degree in astrophysics from Colgate University in 2022, where he conducted multi-wavelength studies of young stellar binaries and developed image processing techniques for Project PANOPTES--a citizen science organization. In 2021, he worked at the Keck Observatory on characterizing static aberrations within the Keck II adaptive optics (AO) system. His current research interests include AO instrumentation and high contrast imaging of exoplanets and protoplanetary disks.

\vspace{1ex}
\noindent Biographies and photographs of the other authors are not available.

\listoffigures

\end{spacing}
\end{document}